\documentclass[aps,prd,amsmath,amssymb,nofootinbib,showpacs,twocolumn]{revtex4}
\usepackage{color}
\usepackage{graphicx}
\usepackage{bm}
\usepackage{bbm}

\newcommand{\Nt}{N_\text{t}}

\newcommand{\vev}[1]{\left\langle #1 \right\rangle}
\newcommand{\ord}{\mathcal{O}}
\newcommand{\abs}[1]{\left \vert #1 \right \vert}
\newcommand{\Sw}{S_{\text{W}}}
\newcommand{\Seff}{S_{\text{eff}}}
\newcommand{\rep}{\mathcal{R}}
\newcommand{\nN}[1]{\left\langle #1 \right\rangle}
\newcommand{\vc}[1]{{\bm{#1}}}
\newcommand{\vcx}{\vc{x}}
\newcommand{\vcy}{\vc{y}}
\newcommand{\trP}{\mathcal{P}}
\newcommand{\coco}{\mathrm{c.c.}}
\newcommand{\Dd}{\mathcal{D}}
\newcommand{\trnsp}{\mathsf{T}}

\DeclareMathOperator{\tr}{tr}

\renewcommand{\Re}{\operatorname{Re}}

\graphicspath{{eps/}}
\begin{document}
\title{Effective Polyakov Loop Dynamics for Finite Temperature $\bm{G_2}$ Gluodynamics}

\author{Bj\"orn H.\ Wellegehausen}
\author{Andreas Wipf}
\author{Christian Wozar}
\thanks{Electronic addresses: \\\texttt{Bjoern.Wellegehausen@uni-jena.de},
\texttt{wipf@tpi.uni-jena.de}
and \texttt{Christian.Wozar@uni-jena.de}}
\affiliation{Theoretisch-Physikalisches Institut,
Friedrich-Schiller-Universit{\"a}t Jena, Max-Wien-Platz 1, 07743
Jena, Germany}


\begin{abstract}
\noindent
Based on the strong coupling expansion we obtain effective 
$3$-dimensional models for the Polyakov loop in finite-temperature $G_2$ 
gluodynamics. The Svetitsky-Jaffe conjecture relates the resulting continuous spin models 
with
$G_2$ gluodynamics near phase transition points.
In the present work we analyse the effective theory in leading
order with the help of a generalised mean field approximation and 
with detailed Monte-Carlo simulations. In addition we derive a 
Potts-type discrete spin model by restricting the characters of the
Polyakov loops to the three extremal points of the fundamental
domain of $G_2$. Both the continuous and discrete effective models 
show a rich phase structure with a ferromagnetic, symmetric and several 
anti-ferromagnetic phases. The phase diagram contains first and second
order transition lines and tricritical points. The modified mean field
predictions compare very well with the results of our simulations.
\end{abstract}
\maketitle


\section{Introduction}
\noindent
For many good reasons $G_2$ gluodynamics has attracted much attention
recently. For example, the $14$-dimensional exceptional group $G_2$ has a trivial centre, 
in contrast to the usually studied $SU(N)$ gauge groups. Indeed
it is the smallest simple and simply connected compact Lie group with this property. 
Thus $G_2$ gluodynamics is useful to better understand the relevance of the 
centre symmetry for confinement \cite{Pepe:2006er}. 
Actually a non-trivial centre is needed in several proposed scenarios for
confinement and hence $G_2$ gluodynamics can be used to test these proposals.
It has been convincingly demonstrated that the theory
shows a first order finite temperature transition without 
order parameter from a confining to a deconfining phase
which can be explained by  centre vortices \cite{Greensite:2006sm}. 
In this context confinement refers to confinement at intermediate scales, where 
a Casimir scaling of string tensions has been reported \cite{Liptak:2008gx}. 
But on large scales, deep in the infrared, strings break due to gluon production and the 
static inter-quark potential becomes flat \cite{Greensite:2003bk}.
Recently it has been demonstrated that chiral symmetry is broken at low
temperatures and is restored at high temperatures at the thermodynamic 
phase transition \cite{Danzer:2008bk}.

$G_2$ gluodynamics has an intriguing connection to $SU(3)$ gauge theory. 
When one couples a scalar field in the $7$-dimensional fundamental
representation to the gauge field one can break the $G_2$ gauge symmetry to the $SU(3)$ gauge symmetry
of strong interaction. With increasing hopping parameter $\kappa$ the resulting 
Yang-Mills-Higgs theory interpolates smoothly between $G_2$ gluodynamics without 
centre symmetry and $SU(3)$ gluodynamics with  $\mathbb{Z}_3$ centre symmetry.
\begin{figure}
\includegraphics{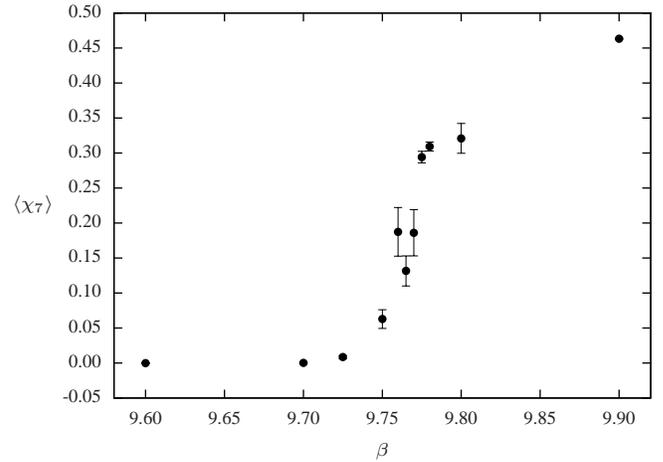}
\caption{Expectation value of the traced Polyakov loop in the
fundamental $7$-dimensional representation in $G_2$ gluodynamics 
on a $16^3\!\times\!6$ lattice as obtained via hybrid Monte-Carlo sampling. The
phase transition is located for the Wilson action at a critical
$\beta_\text{c}\approx 9.765$.}
\label{fig:ymTransition}
\end{figure}
For intermediate values of the hopping parameter the theory mimics $SU(3)$ gauge 
theory with dynamical quarks and the masses of these `quarks' increase 
with increasing hopping parameter. In $G_2$ gluodynamics the Polyakov loop is no longer 
an order parameter in the strict sense.
Despite this fact it still serves as an approximate order 
parameter separating the confined from the deconfined phase (see Fig.~\ref{fig:ymTransition}) 
with a rapid change at the phase transition point. 

According to the conjecture by Svetitsky and Yaffe
\cite{Svetitsky:1982gs,Yaffe:1982qf} the dynamics at the finite temperature
confinement-decon\-finement transition of a $d\!+\!1$-dimensional pure gauge
theory can be described by an effective spin model in $d$ dimensions. Based on
our earlier results on finite-temperature $SU(2)$ and $SU(3)$ gluodynamics
\cite{Dittmann:2003qt,Heinzl:2005xv,Wozar:2007tz,Velytsky:2008bh,Wozar:2008nv}
there are strong indications that the effective models derived and analysed in
the present work are sufficient to accurately describe the dynamics of Polyakov
loops. The direct connection between the effective spin models and  $G_2$
gluodynamics is postponed to a forthcoming publication.

In Sec.~\ref{sect:g2} we review kinematic aspects of
$G_2$ and the main implications for $G_2$ gluodynamics. Afterwards in
Sec.~\ref{sect:strongCoupling} the strong coupling expansion for
the effective Polyakov loop action is explained and in particular the effective 
theory in leading order is introduced. In Sec.~\ref{sect:fundamentalModel} we investigate 
the properties of the effective model first by a classical analysis, then by a modified 
mean field approximation and finally by extensive Monte-Carlo simulations. 
Reducing the continuous spin degrees further to the discrete spins situated at the $3$ 
edges of the fundamental domain of $G_2$ we end up with a deformed Potts-type
spin model whose phase diagram is explored in Sec.~\ref{sect:pottsModel}.


\section{The group $\bm{G_2}$}
\label{sect:g2}
\noindent
$G_2$ is the smallest of the five exceptional simple Lie groups and can be
viewed as a subgroup of $SO(7)$ subject to seven independent cubic constraints for the
$7$-dimensional matrices $g$ representing $SO(7)$ \cite{Holland:2003jy}:
\begin{equation}
\label{eq:g2constraint}
T_{abc} = T_{def}\,g_{da}\,g_{eb}\,g_{fc}.
\end{equation}
Here $T$ is a total antisymmetric tensor given by
\begin{equation}
T_{127} = T_{154} = T_{163} = T_{235} = T_{264} = T_{374} = T_{576} = 1.
\end{equation}
The constraints \eqref{eq:g2constraint} for the group elements reduce the $21$ generators of $SO(7)$ to
$14$ generators of the group $G_2$ with rank $2$. Its fundamental representations are the defining
$7$-dimensional and the adjoint $14$-dimensional representation with Dynkin
labels
\begin{equation}
(7) = [1,0],\quad (14) = [0,1].
\end{equation}
$G_2$ has a trivial centre and its Weyl group is the dihedral group $D_6$ of order $12$.
Additionally $G_2$ is connected to $SU(3)$
through the embedding of $SU(3)$ as a subgroup of $G_2$ according to \cite{Macfarlane:2002hr}
\begin{equation}
G_2/SU(3) \sim SO(7)/SO(6) \sim S_6.
\end{equation}
So when the $S_6$ part of $G_2$ is frozen out\footnote{This is possible when a
fundamental Higgs field is coupled to the gauge field \cite{Pepe:2006er}.} we
end up at $SU(3)$ gauge theory.

\begin{figure}[h]
\includegraphics{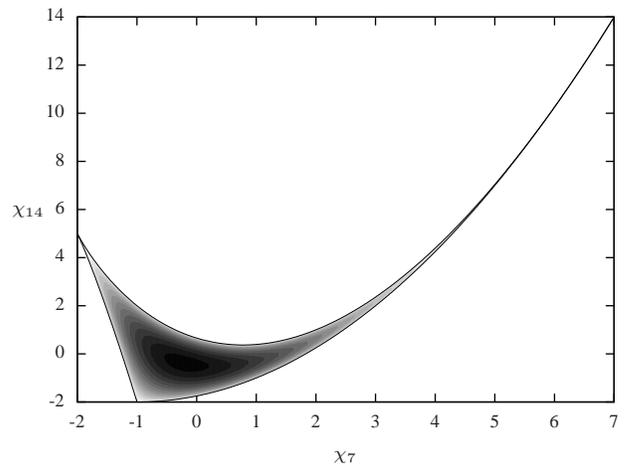}
\caption{Fundamental domain of $G_2$. Darker regions indicate a bigger Haar
measure.}
\label{fig:fundamentalDomain}
\end{figure}

In effective theories for the gauge invariant (traced) Pol\-ya\-kov loops 
in the fundamental representations we are aiming at, only the reduced
Haar measure is needed. Based on \cite{Uhlmann:2006ze} this measure can be 
given for a parametrisation of the conjugacy classes either by angular variables 
or alternatively by the fundamental characters,
\begin{equation}
d\mu \propto J^2d\,\varphi_1\,d\varphi_2 = J\,d\chi_7\,d\chi_{14}.
\end{equation}
The density $J^2$ can be expressed in terms of the fundamental characters,
\begin{equation}
\begin{aligned}
J^2 &= \left(4\chi_7^3-\chi_7^2-2\chi_7-10\chi_7\chi_{14}
+7-10\chi_{14}-\chi_{14}^2\right) \\ &\quad\times
\left(7-\chi_7^2-2\chi_7+4\chi_{14}\right),
\end{aligned}
\end{equation}
where the characters are given in terms of (particularly chosen) angular variables $\varphi_{1,2}$ as
\begin{equation}
\begin{aligned}
\chi_7 &= 1+2\cos(\varphi_1)+2\cos(\varphi_2)+2\cos(\varphi_1+\varphi_2),\\
\chi_{14} &=
2\bigl(1+\cos(\varphi_1) +\cos(\varphi_1-\varphi_2)
+\cos(\varphi_2) \bigr. \\ \bigl. &\;\;+
\cos(\varphi_1+\varphi_2)+\cos(2\varphi_1+\varphi_2) +\cos(\varphi_1+2\varphi_2) \bigr).
\end{aligned}
\end{equation}
The boundary of the fundamental domain is determined by $J=0$ and thus is parametrised by 
the three curves (see Fig.~\ref{fig:fundamentalDomain})
\begin{equation}
\begin{aligned}
\chi_{14} &= \frac{1}{4}(\chi_7+1)^2-2,\\
\chi_{14} &= -5(\chi_7+1)\pm 2(\chi_7+2)^{3/2}.
\end{aligned}
\end{equation}
Note that the reduced $G_2$-Haar measure is maximal not at the origin
but for $(\chi_7,\chi_{14})=(-1/5,-2/5)$. The fundamental domain has no 
symmetries at all and this expresses the fact that the centre of
$G_2$ is  trivial.

\subsection*{Representation theory and implications for confinement}
\noindent
In the pioneering work \cite{Holland:2003jy} the confining properties  of $G_2$
have been discussed and compared to those of $SU(3)$. Quarks and anti-quarks 
in $SU(3)$ transform under the fundamental  representations $3$ and $\bar 3$
such that their charges can only be screened by particles with non-vanishing $3$-ality,
especially \emph{not by gluons}. This explains why in the confining phase of 
$SU(3)$ gluodynamics the static inter-quark potential is linearly rising up to arbitrary 
long distances. As a consequence the free energy of a single quark gets infinite and
the Polyakov loop expectation value vanishes. Hence the Polyakov loop discriminates
the confining from the deconfining phase and at the same time serves as order 
parameter for the $\mathbb{Z}_3$ centre symmetry.

To better understand $G_2$ gluodynamics we recall the decomposition of
tensor products into irreducible representations,
\begin{equation}
\label{eq:representationsG2}
\begin{aligned}
(7) \otimes (7)&=(1) \oplus (7) \oplus (14) \oplus (27)\\ 
(7) \otimes (7) \otimes (7)&=(1) \oplus 4 \cdot (7) \oplus 2 \cdot (14) \oplus 3 \cdot (27) \\
&\quad \oplus 2 \cdot (64) \oplus (77)\\
(14) \otimes (14)&=(1) \oplus (14) \oplus (27) \oplus (77) \oplus (77') \\
(14) \otimes (14)\otimes (14)&=(1)\oplus (7) \oplus 5\cdot (14)\oplus 
3\cdot(27)\oplus\dotsb
\end{aligned}
\end{equation}
with Dynkin labels
\begin{equation}
\begin{aligned}
(1) &= [0,0],& (27) &= [2,0],& (64) &= [1,1],\\ (77)&=
[3,0],& (77') &= [0,2].
\end{aligned}
\end{equation}
The quarks in $G_2$ transform under the $7$-dimensional fundamental representation, gluons
under the $14$-dimensional fundamental (and at the same time adjoint) representation. From
\eqref{eq:representationsG2} we see that similarly as in $SU(3)$ two or three quarks
can build a colour singlet (meson or baryon, respectively). In $G_2$ gluodynamics three
centre-blind dynamical gluons can screen the colour charge of a single quark,
\begin{equation}
(7) \otimes (14) \otimes (14) \otimes (14)=(1) \oplus \dotsb.
\end{equation}
Thus the flux tube between two static quarks can break due to gluon production
and the Polyakov loop does not vanish even in the
confining phase \cite{Pepe:2006er}. This shows that the Polyakov loop can at
best be an approximate order parameter (see Fig.~\ref{fig:ymTransition}) which 
changes rapidly at the phase transition and is small (but non-zero) in the confining phase. 
To characterise confinement we can no longer refer to a non-vanishing
asymptotic string tension and vanishing Polyakov loop. Instead we define
confinement as the absence of free colour charges in the physical spectrum. 
In the confining phase the inter-quark potential rises linearly
at intermediate scales \cite{Greensite:2006sm,Liptak:2008gx}.


\section{Effective theories and the strong coupling expansion}
\label{sect:strongCoupling}
\noindent
Based on a conjecture  relating
finite temperature $SU(N)$ gluodynamics in $d\!+\!1$ dimensions at the critical
point with a $\mathbb{Z}_N$ spin model in $d$ dimensions \cite{Svetitsky:1982gs,Yaffe:1982qf}, 
there have been  extended studies to compare correlation functions of both systems for $SU(2)$
\cite{Dittmann:2003qt,Heinzl:2005xv,Velytsky:2008bh} and $SU(3)$ gluodynamics
\cite{Wozar:2007tz,Wozar:2008nv}, either by using Schwinger-Dyson equations 
or demon methods \cite{Creutz:1983ra,Hasenbusch:1994ne}.
The strong coupling expansion for the distribution of the
inhomogeneous Polyakov loops was taken as ansatz for the (exponentiated)
effective Polyakov-loop action. This way effective models for
$SU(3)$ gluodynamics have been derived in \cite{Wozar:2007tz}. Here
we sketch how one arrives at the analogous results for $G_2$ and obtain
the effective continuous spin model in leading order.

Starting with the lattice Wilson action
\begin{equation}
 \Sw = \beta \sum_{\square} \left( 1-  \frac{1}{N_\text{C}} \Re \tr
 U_{\square} \right),\quad \beta = \frac{6}{a^4g^2}
\end{equation}
a \emph{strong coupling expansion} (for small
$\beta$) is performed to arrive at an effective theory for the local
Polyakov loops. To do that one inserts a group valued delta
function into the path integral,
\begin{equation}
\begin{aligned}
Z &= \int \Dd U \exp(-\Sw[U])\\
&= \int \Dd P \int \Dd U\, \delta\left(P_\vcx,\prod_{\tau=1}^{\Nt}
U_{\tau,\vcx}\right) \exp(-\Sw[U])
\\ &\equiv \int \Dd \trP\, \exp(-\Seff[\trP]).
\end{aligned}
\end{equation}
Here $\Dd P$ denotes the product of reduced Haar measures on the sites
of the \emph{spatial lattice}. We do not need the full Haar measure of $G_2$ 
since the effective action $\Seff$ only depends 
on the gauge invariant content of the local Polyakov loop.

In compact form the strong coupling
expansion is then given by
\begin{equation}
\begin{aligned}
  \Seff[\trP] &= \sum_r \sum_{\rep_1 \dotsc  \rep_r} \sum_{\ell_1 \dotsc \ell_r}
  c_{\rep_1 \dotsc \rep_r}^{\ell_1 \dotsc  \ell_r}(\beta)  \prod_{i=1}^r
  S_{\rep_i, \ell_i} \\&= \sum_i \lambda_i S_i
\end{aligned}
\end{equation}
with the basic building blocks
\begin{equation}
  S_{\rep, \ell} \equiv \sum \limits_{\vcx,\vcy}\chi_\rep (\trP_\vcx) \chi_\rep^* (\trP_{\vcy}) + \coco
  ,  \quad \abs{\vcx-\vcy}=\ell.
\end{equation}
Here $r$ counts the number of link operators contributing at
each order. The coefficients $c_{\rep_1 \dotsc \rep_r}^{\ell_1
\dotsc \ell_r}$ couple the operators
$S_{\rep_i,\ell_i}$ sitting at sites $\vcx_i$ and $\vcy_i$ separated 
by distance $\ell_i$ 
in representation $\rep_i$. The effective action hence describes a \emph{network of link operators} that are collected into (possibly disconnected)
`polymers' contributing with `weight' $c_{\rep_1 \dotsc
\rep_r}^{\ell_1 \dotsc \ell_r}$. The resulting `operators'
(Polyakov loop monomials) are \emph{dimensionless} and there is
\emph{no natural} ordering scheme at hand. Our chosen truncation is based
on the strong coupling expansion in powers of $\beta$ which is 
closely related to the dimension of the corresponding group 
representations and the distance across which the Polyakov loops 
are coupled. In the strong coupling expansion truncated 
at $\ord(\beta^{k\Nt})$ one has $r \le k$ and the additional 
restriction $|\rep_1| + \dotsb + |\rep_r| < k$ with $|\rep| \equiv p_1+p_2 $
for a given representation $\rep$ of $G_2$ with Dynkin labels
$[p_1,p_2]$. The leading order terms only contain
interactions between nearest neighbours $\nN{\vcx\vcy}$ and 
the two fundamental representations.

For $SU(3)$ the characters of the two fundamental representations are 
complex conjugate of each other such that the effective 
Polyakov loop action contains just one term in leading order. 
In $G_2$ this situation  changes and we find two
independent contributions in leading order. We refer to the
corresponding model containing the two fundamental representations
as \emph{fundamental model}. Its action is explicitly given by
\begin{equation}
\label{eq:fundamentalEffectiveAction}
S_\text{{eff}}=\lambda_7 \underbrace{\sum \limits_{\nN{\vcx\vcy}}
\chi_7(\trP_\vcx) \chi_7(\trP_\vcy)}_{S_{7}} +
\lambda_{14} \underbrace{\sum \limits_{\nN{\vcx\vcy}} \chi_{14}(\trP_\vcx)
\chi_{14}(\trP_\vcy)}_{S_{14}},
\end{equation}
where the couplings $\lambda_7$ and $\lambda_{14}$ are indexed by the dimension 
of the involved representation. In next-to leading order
there exist six additional terms with nearest neighbour interactions.
Their explicit forms are dictated by representation theory \eqref{eq:representationsG2} 
\begin{equation}
\begin{aligned}
S_{27}&=\sum \limits_{\nN{\vcx\vcy}} \chi_{27}(\trP_\vcx)
\chi_{27}(\trP_\vcy), \\
S_{77'}&=\sum \limits_{\nN{\vcx\vcy}}
\chi_{77'}(\trP_\vcx) \chi_{77'}(\trP_\vcy), \\
S_{64}&=\sum
\limits_{\nN{\vcx\vcy}} \chi_{64}(\trP_\vcx) \chi_{64}(\trP_\vcy), \\
S_{7,7}&=\sum \limits_{\nN{\vcx\vcy}} \left(\chi_7(\trP_\vcx)
\chi_7(\trP_\vcy) \right)^2, \\
S_{14,14}&=\sum \limits_{\nN{\vcx\vcy}}
\left(\chi_{14}(\trP_\vcx) \chi_{14}(\trP_\vcy) \right)^2, \\
S_{7,14}&=\sum \limits_{\nN{\vcx\vcy}} \chi_7(\trP_\vcx)\chi_7(\trP_\vcy)
\chi_{14}(\trP_\vcx)  \chi_{14}(\trP_\vcy).
\end{aligned}
\end{equation}
It the remainder of this work we shall neglect the next-lo leading order
terms and concentrate on the fundamental model
\eqref{eq:fundamentalEffectiveAction}.


\section{The fundamental model}
\label{sect:fundamentalModel}
\noindent
For the fundamental effective model \eqref{eq:fundamentalEffectiveAction} we
shall localise the symmetric, ferromagnetic and anti-ferromagnetic phases
with coexistence lines in order to find
the region in the space of couplings $\lambda_7,\lambda_{14}$
where a connection to $G_2$ gluodynamics can be established.

\subsection{Classical analysis}
\begin{figure}[h]
\includegraphics{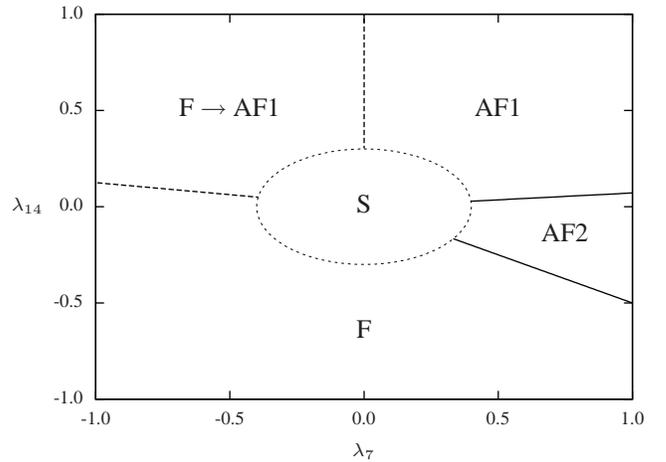}
\caption{The classical phase diagram of the fundamental effective
$G_2$ model. In addition to the calculated ferromagnetic
and anti-ferromagnetic phases we expect a symmetric phase 
for weak couplings.}
\label{fig:effectiveClassical}
\end{figure}

\noindent
For strong couplings the fluctuations of the Polyakov loops are suppressed
and the spin system behaves almost classically. Thus for large  $\vert\lambda_7\vert$ and 
$\vert\lambda_{14}\vert$ we may compute the phase diagram by minimising the 
classical action. Anticipating that there are anti-ferromagnetic phases we 
introduce the odd and even sublattices
\begin{equation}
\begin{aligned}
\Lambda_\text{o}&=\left \lbrace \vcx \,\vert\,x_1+x_2+x_3\text{ odd}
\right \rbrace\quad\text{and}\\ 
\Lambda_\text{e}&=\left \lbrace \vcx\,\vert\,
x_1+x_2+x_3\text{ even}\right \rbrace.
\end{aligned}
\end{equation}
On each sublattice the Polyakov loop is assumed to 
have a \emph{constant} value and the two values are denoted by 
$\trP_\text{o}$ and $\trP_\text{e}$, respectively. We denote the corresponding characters
in the fundamental domain of $G_2$ by
\begin{equation}
\begin{aligned}
\vc{\chi}_\text{e}&=\begin{pmatrix}\chi_{7,\text{e}}\\ \chi_{14,\text{e}}\end{pmatrix}=
\begin{pmatrix}\chi_{7}\\ \chi_{14}\end{pmatrix}(\trP_\text{e}) \quad\text{and} \\
\vc{\chi}_\text{o}&=\begin{pmatrix}\chi_{7,\text{o}}\\ \chi_{14,\text{o}}\end{pmatrix}
=\begin{pmatrix}\chi_{7}\\ \chi_{14}\end{pmatrix}(\trP_\text{o}).
\end{aligned}
\end{equation}
With this assumption and notation the action of the fundamental model 
\eqref{eq:fundamentalEffectiveAction}
reads
\begin{equation}
\Seff=V
\vc{\chi}^\trnsp_\text{e}K\vc{\chi}_\text{o},\quad\text{with}\quad 
K=3\begin{pmatrix}\lambda_7&0\\0&\lambda_{14}\end{pmatrix}.\label{actioncl}
\end{equation}
To localise the different phases we may assume that the Polyakov loop on 
one sublattice, say $\Lambda_\text{o}$, is equal to the group-identity with maximal
characters, $\chi_{7,\text{o}}=7$ and $\chi_{14,\text{o}}=14$.
For given couplings $\lambda_7$ and $\lambda_{14}$ the corresponding
thermodynamic phase is then determined by that Polyakov loop on $\Lambda_\text{e}$ 
for which the linear function $7\lambda_7\chi_{7,\text{e}}+14\lambda_{14}\chi_{14,\text{e}}$
is minimal. If the minimising characters are the same on 
both sublattices then the phase is ferromagnetic,
else it is anti-ferromagnetic. The minimum of the linear function is
attained for $\vc{\chi}_\text{e}$ on one of the three corners of the fundamental 
domain in Fig.~\ref{fig:fundamentalDomain} or on the curve connecting the 
corners $(-1,-2)^\trnsp$  and $(7,14)^\trnsp$. Depending on the sign of
$\lambda_7$ and the slope $\xi\equiv\lambda_{14}/\lambda_7$ one finds the following phases:
\begin{itemize}
\item  For $\lambda_7>0$ and $\xi<-1/2$ or for $\lambda_{7}<0$ and
$\xi>-1/8$ we find the \emph{ferromagnetic phase} F with
$\vc{\chi}^\trnsp_\text{e}=(7,14)$.
\item For $\lambda_{7}>0$ and $\xi>1/14$
we find a \emph{anti-ferromagnetic phase} AF1 with
$\vc{\chi}^\trnsp_\text{e}=(-1,-2)$.
\item For $\lambda_7>0$ and $-1/2<\xi<1/14$ we find a second 
\emph{anti-ferromagnetic phase} AF2 with $\vc{\chi}^\trnsp_\text{e}=(-2,5)$.
\item For $\lambda_{7}<0$ and $\xi<-1/8$ the characters
$\vc{\chi}^\trnsp_\text{e}=(-1-1/\xi, 1/(2\xi)^2-2)$ change continuously from $(7,14)$ to $(-1,-2)$ 
along the connecting boundary curve of the fundamental domain. This \emph{transition
phase} is denoted by F~$\to$~AF1.
\end{itemize}
The phase portrait is depicted in Fig.~\ref{fig:effectiveClassical}
where we also included the expected symmetric phase for weak 
couplings. Since in a symmetric phase entropy wins over energy
it cannot be seen in any classical analysis.

\subsection{Mean field analysis}
\noindent
The classical analysis is refined by a modified mean field approximation 
in which one allows for inhomogeneous mean fields.  First we
recall the main aspects of the method. Here we are interested 
in expectation values of the form
\begin{equation}
\begin{aligned}
\vev{A} &= \frac{1}{Z}\int \Dd \trP \exp(-\Seff[\trP]) A[\trP],\\
\Dd \trP &\equiv \prod_\vcx d\mu(\trP_\vcx),
\end{aligned}
\end{equation}
where the gauge invariant function $A(\trP)$ depends on the Polyakov
loop via the the fundamental characters $\chi_7(\trP)$ 
and $\chi_{14}(\trP)$ and $d\mu$
is the reduced Haar measure of $G_2$. The equilibrium probability measure $\Dd \trP
\exp(-\Seff)/Z$ is the unique minimum to the variational problem
\begin{equation}
\label{eq:variationalProblem}
\inf_p \vev{\Seff+\ln p}_p,
\end{equation}
where the $p$-indexed expectation value is calculated with the 
integration measure $p[\trP]\,\Dd\trP$, whose probability 
density $p$ is to be varied. Expectation values of observables can 
then be computed as
\begin{equation}
\vev{A}_p =\int \Dd \trP \,p[\trP] A[\trP].
\end{equation}
In this scope a Monte-Carlo simulation is just the approximation of 
the probability density $p[\trP]\propto\exp(-\Seff[\trP])$ with a finite 
set of configurations which give 
$p[\trP]\approx N_\text{MC}^{-1}\times\sum_{t=1}^{N_\text{MC}}
\delta(\trP-\trP_t)$, where $\trP_t$ is the configuration in the $t$'th MC step and $N_\text{MC}$ 
is the number of MC steps.

In a variational approach the mean field approximation amounts to the restriction 
of the admissible densities $p$ to product form
\begin{equation}
p[\trP]\rightarrow \prod_\vcx p_\vcx(\trP_\vcx).
\end{equation} 
Then expectation values factorise and the computation can be done site by site.
Due to the translational invariance of the action one may believe, that the
minimising density is translational invariant, $p_\vcx(\trP)=p(\trP)$. However,
this assumption is only justified for the symmetric and ferromagnetic phases with 
constant mean fields.

\begin{figure}[t]
\includegraphics{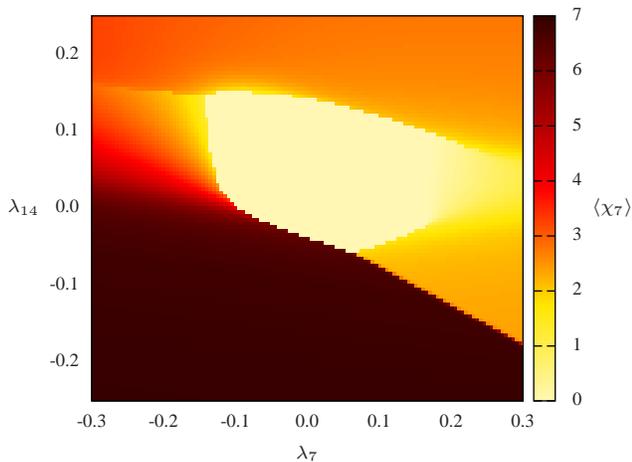}
\caption{Polyakov loop $\langle\chi_7\rangle$ of the 
\emph{fundamental} $G_2$ spin model in
mean field approximation.}
\label{fig:effectiveMeanfield}
\end{figure}

Anticipating the existence of additional anti-ferromagnetic phases we 
partition the lattice into its even and odd sub-lattices, as we did in the classical
analysis, and allow for different densities on the sublattices,
\begin{equation}
p_\vcx(\trP_\vcx) = \begin{cases}
           p_\text{e}(\trP_\vcx)&: \vcx\in\Lambda_\text{e}\\
           p_\text{o}(\trP_\vcx)&: \vcx\in\Lambda_\text{o}.
           \end{cases}
\end{equation}
The classical analysis is then recovered by allowing only $\delta$-type
point-measures for $p_{\text{e},\text{o}}$. In the modified mean-field analysis
we allow for all $p_{\text{e},\text{o}}$ in the variational principle with prescribed
mean fields $\bar{\vc\chi}_\text{e}$ and $\bar{\vc\chi}_\text{o}$ on the even and odd sublattices. The effective potential $u$ is then obtained by computing
\begin{equation}
u(\bar{\vc{\chi}}_\text{e},\bar{\vc{\chi}}_\text{o}) =\frac{1}{V}\inf_p \vev{\Seff+\ln p}_p,
\end{equation}
subject to the following four constraints for the admitted densities $p_\text{e}$ and $p_\text{o}$:
\begin{equation}
\vev{\vc\chi}_{\text{e},\text{o}}=\bar{\vc\chi}_{\text{e},\text{o}}.
\label{constraints}
\end{equation}
The one-site expectation values $\vev{\dots}_{\text{e},\text{o}}$ are calculated with
$p_{\text{e},\text{o}}\,d\mu$. To actually compute the minimising densities one needs the
expectation value of the action and  entropy, given by
\begin{equation}
\begin{aligned}
\vev{\Seff}_p &= 
V\bar{\vc\chi}^\trnsp_\text{e} K\bar{\vc\chi}_\text{o} \\
\vev{\ln p}_p &= \frac{V}{2} \vev{\ln p_\text{e}}_\text{e}+\frac{V}{2}\vev{\ln p_\text{o}}_\text{o},
\end{aligned}
\end{equation}
where $K$ is the matrix given in \eqref{actioncl}.
On each sublattice the variational problem is solved by a 
density $p\propto \exp\bigl(\vc{j}\cdot\vc{\chi}(\trP)\bigr)$
with two Lagrangian multipliers $\vc{j} =(j_7,j_{14})$. 
The four multipliers are determined by the four constraints in
\eqref{constraints}.
Using this solution for the densities in the variational principle determines the 
effective potential as function of the prescribed mean fields. The expectation
values of the characters on the two sublattices minimise the effective
potential. These minimas solve the following system of coupled gap equations
\begin{equation}
\begin{aligned}
K\bar{\vc\chi}_\text{e}&=-\frac{\partial w(\bar{\vc \chi}_\text{o})}{\partial \bar{\vc\chi}_\text{o}},\quad
K\bar{\vc\chi}_\text{o}=-\frac{\partial w(\bar{\vc \chi}_\text{e})}{\partial \bar{\vc\chi}_\text{e}},\\
w(\bar{\vc\chi})&=\ln\int d\mu(\trP)\, e^{-\bar{\vc\chi}^\trnsp
K\vc\chi(\trP)}.
\end{aligned}
\end{equation}
We have calculated the expectation values of $\chi_7$ and $\chi_{14}$
on both sublattices as functions of
the couplings on a $120\!\times\!100$ grid in the rectangle
\begin{equation}
-0.3\leq \lambda_7\leq 0.3\quad\text{and}\quad
-0.25\leq\lambda_{14}\leq 0.25.
\label{grid}
\end{equation}
The contour plot of the expectation value
\begin{equation}
\langle \chi_7\rangle=\frac{1}{2}\left(\langle\chi_{7,\text{e}}\rangle
+\langle\chi_{7,\text{o}}\rangle\right),\label{magnetization}
\end{equation}
called \emph{magnetisation},
is depicted in Fig.~\ref{fig:effectiveMeanfield}. As expected, for weak couplings we find
a symmetric phase with vanishing magnetisation in the centre of the phase diagram. 
On the lower left, for negative couplings, we find the ferromagnetic phase
with $\langle\chi_{7,\text{e}}\rangle=\langle\chi_{7,\text{o}}\rangle\approx 7$ 
or equivalently with a typical $\trP_\vcx$ near the identity. For an unambiguous
identification of the phases one needs the expectation values
of both $\chi_7$ and $\chi_{14}$ on both sublattices. We have calculated
these four expectation values for the fundamental model on
a grid in the space of coupling constants with extensive Monte Carlo simulations.
Since the numerical simulations and mean field approximation yield almost
identical results we defer the detailed discussion of
phase portrait, and in particular the localisation of the various 
anti-ferromagnetic phases for positive couplings,  to the following section.

\subsection{Monte-Carlo results} 
\begin{figure}[h]
\includegraphics{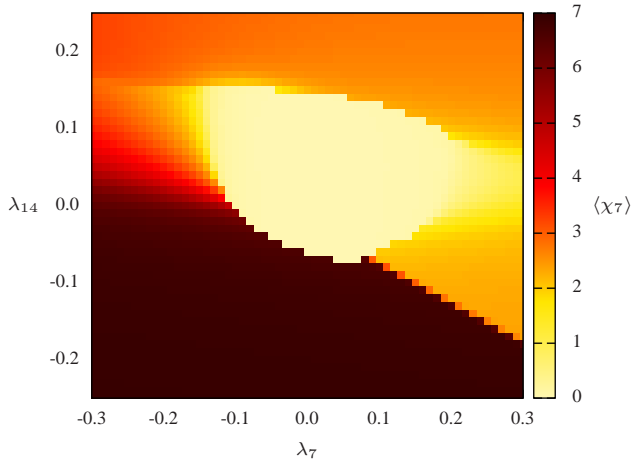}
\caption{Polyakov loop $\langle\chi_7\rangle$ of the 
fundamental model obtained via MC simulation on an $8^3$ lattice.}
\label{fig:effectivePolyakov}
\end{figure}

\begin{figure}[t]
\includegraphics{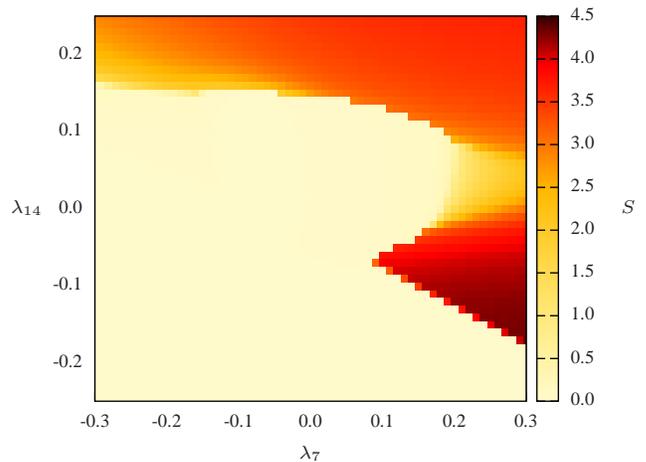}
\caption{Staggered magnetisation $S$ of the fundamental
model obtained via MC simulation on an $8^3$ lattice.}
\label{fig:effectiveStaggered}
\end{figure}

\begin{figure}[h]
\includegraphics{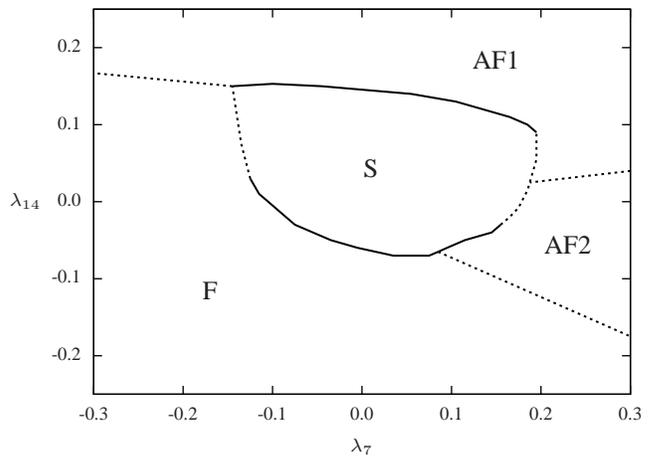}
\caption{Phase diagram of the fundamental effective model. The orders of phase
transitions are indicated with full lines (first order) and dotted lines
(second order/crossover).}
\label{fig:effectivePhases}
\end{figure}

\noindent
We performed our Monte-Carlo simulations with about $10\,000$ 
samples for every point on a $60\!\times\!50$ grid inside 
the rectangle \eqref{grid} in the space of coupling constants.
Two neighbouring points on this grid are separated by $0.01$.
First we calculated the magnetisation $\langle\chi_7\rangle$ and
the resulting phase portrait is depicted in Fig.~\ref{fig:effectivePolyakov}. 
It looks very similar to the portrait calculated in the mean field approximation,
see Fig.~\ref{fig:effectiveMeanfield}.
For weak couplings entropy wins over energy and the product of
Haar measures of the Polyakov loops become relevant.
In order to unambiguously identify the anti-ferromagnetic
phases for positive couplings we again subdivided the lattice in
the even and odd sublattice, $\Lambda=\Lambda_\text{e}\cup\Lambda_\text{o}$, 
and measured the \emph{staggered magnetisation}
\begin{equation}
S=\frac{1}{2}\langle\abs{\chi_{7,\text{e}}-\chi_{7,\text{o}}}\rangle.\label{staggered}
\end{equation}
The corresponding contour plot is exhibited in Fig.~\ref{fig:effectiveStaggered}.
On top and on the right of the plot the staggered magnetisation gets large
and we identify this region as belonging to anti-ferromagnetic phases.
For large absolute values of $\lambda_7,\,\lambda_{14}$ action (energy)
dominates over entropy and this explains why the simulation results agree well 
with the classical analysis for strong couplings: all phases but the \emph{transition phase} 
F~$\to$~AF1 are already  visible in the classical phase diagram
in Fig.~\ref{fig:effectiveClassical}.
In the Monte-Carlo simulation an additional  ``symmetric phase'' with vanishing
Polyakov loop and vanishing staggered magnetisation appears for
weak couplings, in complete agreement with our mean field
analysis. The resulting phase diagram with one symmetric,
one ferromagnetic and two anti-ferromagnetic phases is depicted 
in Fig.~\ref{fig:effectivePhases}.

\begin{figure}[t]
\includegraphics{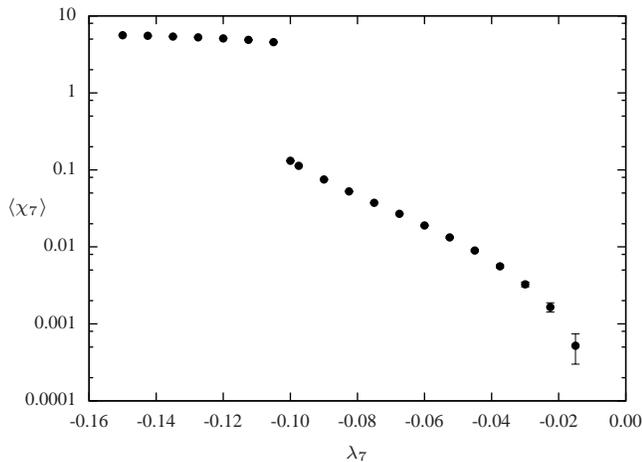}
\caption{Polyakov loop $\vev{\chi_7}$ of the fundamental effective model with
coupling $\lambda_{14}=0$ obtained via MC simulation on an $8^3$ lattice.}
\label{fig:symmetricNonVanishing}
\end{figure}

Eventually the finite temperature phase transition in $G_2$ gluodynamics 
will correspond to a transition between the symmetric and the
ferromagnetic phase in the effective spin model. The dependence of 
the effective couplings $\lambda_7,\lambda_{14}$ on the Yang-Mills coupling
$g$ can be calculated with the help of powerful inverse Monte-Carlo 
techniques \cite{Wozar:2007tz,Wozar:2008nv}. This will be done in a 
forthcoming publication.
However, we anticipate that the confinement-deconfinement phase
transition in $G_2$ gluodynamics will happen near the critical point
$\lambda_{14,\text{c}}=0,\;\lambda_{7,\text{c}}\approx -0.0975(75)$ of the
fundamental model. Thus we have plotted the magnetisation in the vicinity of this
first order transition from the ferromagnetic to the symmetric phase in 
Fig.~\ref{fig:symmetricNonVanishing}.
Note that even in the ``symmetric'' phase we find a non-zero magnetisation
$\langle \chi_7\rangle$ which jumps at the critical coupling
$\lambda_{7,\text{c}}$. This parallels the jump of the Polyakov loop in
$G_2$ gluodynamics, see Fig.~\ref{fig:ymTransition}.

\begin{figure}[t]
\includegraphics{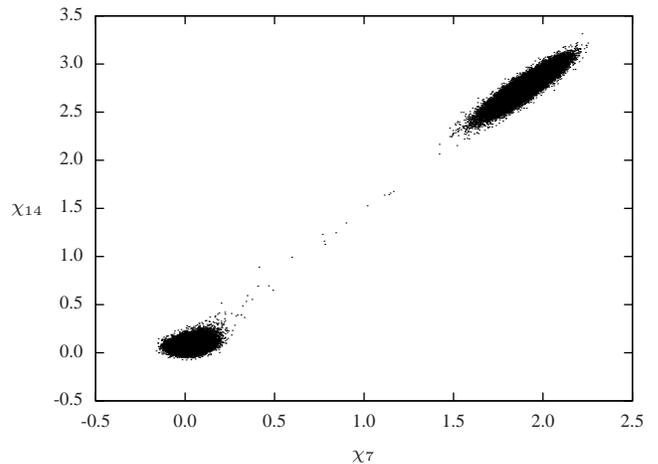}
\caption{Distribution of $\chi_7$ and $\chi_{14}$
in the fundamental domain of $G_2$ at $\lambda_7=0$ and $\lambda_{14}=0.1446$.}
\label{fig:Histo_1}
\end{figure}

The phase diagram in Fig.~\ref{fig:effectivePhases} contains lines of second and
first order transitions and $3$ triple points. The full lines belong to first order
and the dotted lines to second order transitions. Note that we may pass from
the symmetric to the ferromagnetic phase via a first or via a second order
transition. The transitions from the ferromagnetic to the anti-ferromagnetic
phases AF1 and AF2 and between the anti-ferromagnetic phases are always
of second order. In order to determine the orders of the transitions we calculated more than $30$ 
histograms for the Polyakov loop distribution near the various phase transition curves
and the changes of various `order parameters' when one crosses the transition
lines. A typical scatter plot is depicted in Fig.~\ref{fig:Histo_1}. It shows
the distribution of $\chi_7$ at a transition from the symmetric to the anti-ferromagnetic
phase AF1 with critical couplings $\lambda_{7,\text{c}}=0$ and
$\lambda_{14,\text{c}}=0.1446$. Without further analysis it is already clear
that we are dealing with a first order phase transition.

The following Fig.~\ref{fig:PolyakovS0_1} shows the behaviour of the magnetisations
$\langle\chi_7\rangle$ and $\langle\chi_{14}\rangle$ near the transition from the
symmetric to the ferromagnetic phase, which happens for $\lambda_{14}=0.13$ and
$\lambda_{7}$ between $-0.18$ and $-0.12$. Both expectation values vary continuously
during the transition and this already suggests that the transition is of second
order. This conclusion is further substantiated by the corresponding histograms for the distributions of 
$\chi_7$ and $\chi_{14}$.

\begin{figure}[t]
\includegraphics{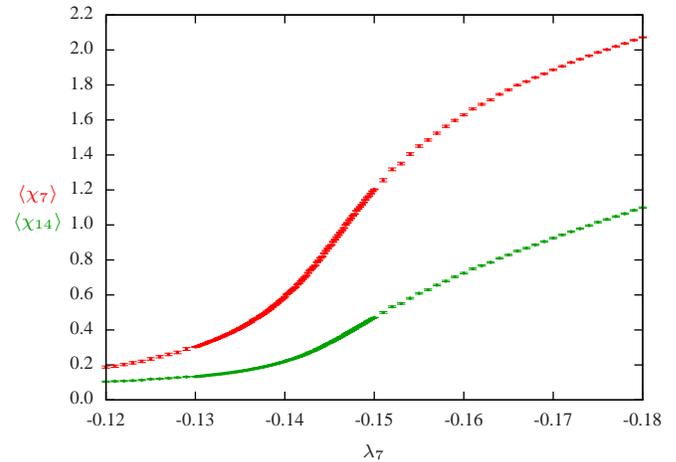}
\caption{Magnetisations $\langle\chi_7\rangle$ (upper curve) and
$\langle\chi_{14}\rangle$ (lower curve) for various $\lambda_7$ at
$\lambda_{14}=0.13$.}
\label{fig:PolyakovS0_1}
\end{figure}


\section{The $\bm{G_2}$ Potts model}
\label{sect:pottsModel}
\noindent
After having collected sufficient information to reconstruct the full phase
diagram of the fundamental continuous spin model with two 
effective couplings we now truncate the degrees of freedom further
to arrive at a discrete spin model. In the case of the
well-studied $SU(3)$ Polyakov loop models one projects a Polyakov loop 
$\trP_\vcx$ onto the closest centre elements of $SU(3)$ and arrives 
at a $\mathbb{Z}_3$ Potts model with action (energy) given by
\begin{equation}
S_N = -\beta \sum_{\langle \vcx\vcy\rangle} \delta(\sigma_\vcx,\sigma_\vcy),\quad \sigma_\vcx\in \mathbb{Z}_N.
\end{equation}
The continuous and discrete models have coinciding critical exponents at
the second order anti-ferromagnetic phase transition and similar phase structures
\cite{Wozar:2006fi}. Motivated by these earlier successes we perform a similar
reduction of the fundamental $G_2$ spin model and arrive at a discrete 
Potts-like $G_2$ spin model.

\begin{figure}[t]
\includegraphics{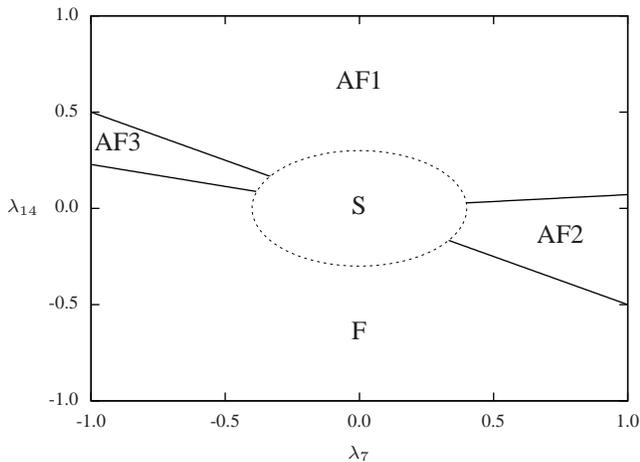}
\caption{The classical phase diagram of the discrete $G_2$ Potts model.}
\label{fig:SpinClassical}
\end{figure}

By projecting the values of $\trP_\vcx$ to the three group elements with 
characters $\chi_7,\chi_{14}$ lying at the extremal points of the 
fundamental domain in Fig.~\ref{fig:fundamentalDomain},
we arrive at a model for the tree spins
\begin{equation}
\sigma_\vcx\in\left\{
\begin{pmatrix}7\\14\end{pmatrix},
\begin{pmatrix}-1\\-2\end{pmatrix},
\begin{pmatrix}-2\\5\end{pmatrix}
\right\}
\end{equation}
with nearest neighbour interaction determined by the Potts-type action
\begin{equation}
S_\text{Potts} = \sum_{\langle \vcx, \vcy\rangle}\sigma_\vcx^\trnsp
\begin{pmatrix}
\lambda_7 & 0\\ 0 &
\lambda_{14}
\end{pmatrix}
\sigma_\vcy\,.
\end{equation}
As expected, the classical phase diagram of the Potts-type model 
with discrete spins is similar to the diagram of the fundamental model with continuous spins. 
Depending on the sign of $\lambda_7$ and the slope $\xi=\lambda_{14}/\lambda_7$
we find the following phases and phase transition lines:
\begin{itemize}
\item For $\lambda_7>0$ and $\xi>-1/2$ and for $\lambda_7<0$ and
$\xi>-47/206$ we find the \emph{ferromagnetic phase}
F with $\vc{\chi}_\text{e}=\vc{\chi}_\text{o}=(7,14)^\trnsp$.
\item For $\lambda_7<0$ and $-1/2<\xi<-47/206$ we find the \emph{anti-ferromagnetic
phase} AF3 with $\vc{\chi}_\text{e}=(-1,-2)^\trnsp$ and $\vc{\chi}_\text{o}=(-2,5)^\trnsp$.
\item  For $\lambda_7>0$ and $-1/2<\xi<1/14$ we find the \emph{anti-ferromagnetic phase} AF2 wit 
$\vc{\chi}_\text{e}=(-2,5)^\trnsp$ and $\vc{\chi}_\text{o}=(7,14)^\trnsp$.
\item For $\lambda_7<0$ and $\xi<-1/2$ and for $\lambda_7>0$ and $\xi>1/14$ we find the \emph{anti-ferromagnetic phase} AF1 
with $\vc{\chi}_\text{e}=(-1,-2)^\trnsp$ and $\vc{\chi}_\text{o}=(7,14)^\trnsp$.
\end{itemize}
The phase portrait is depicted in Fig.~\ref{fig:SpinClassical}, where we have 
inserted by hand the expected symmetric phase for weak couplings.
A striking difference between the diagrams in Fig.~\ref{fig:effectiveClassical} 
and in Fig.~\ref{fig:SpinClassical} is the absence of the ``transition phase'' 
F~$\to$~AF1 in the discrete model for which this phase does not exist by
construction. Instead we find a third anti-ferromagnetic phase denoted by AF3 in Fig.~\ref{fig:SpinClassical}.
In addition, in the symmetric phase of the continuous spin model $\vev{\chi_7}\approx 0$
and in the symmetric phase of the discrete spin model $\vev{\chi_7}\approx 4/3$.

\begin{figure}[t]
\includegraphics{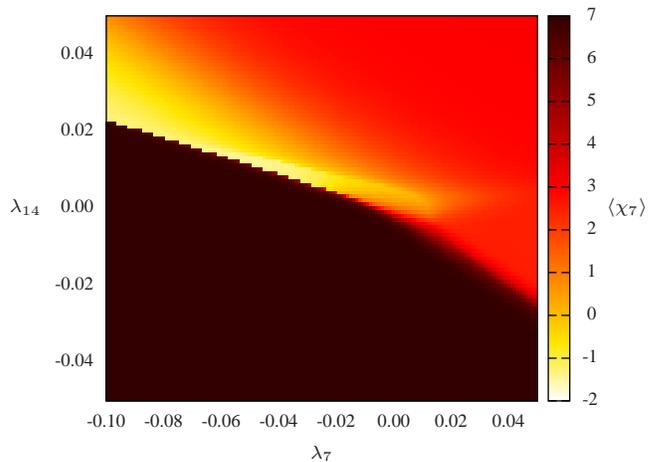}
\caption{Magnetisation $\vev{\chi_7}$ of the discrete $G_2$ Potts model in
mean field approximation.}
\label{fig:SpinMeanfield}
\end{figure}

Similarly as for the continuous model we calculated the phase diagram of the discrete model 
with the help of the modified  mean field approximation. The contour plot for the magnetisation
is depicted in Fig.~\ref{fig:SpinMeanfield}.
In the lower part of the plot we can see the ferromagnetic phase for which the Polyakov
loops on both sublattices $\Lambda_\text{e}$ and $\Lambda_\text{o}$ are equal to the identity
with very high probability.

\begin{figure}[htb]
\includegraphics{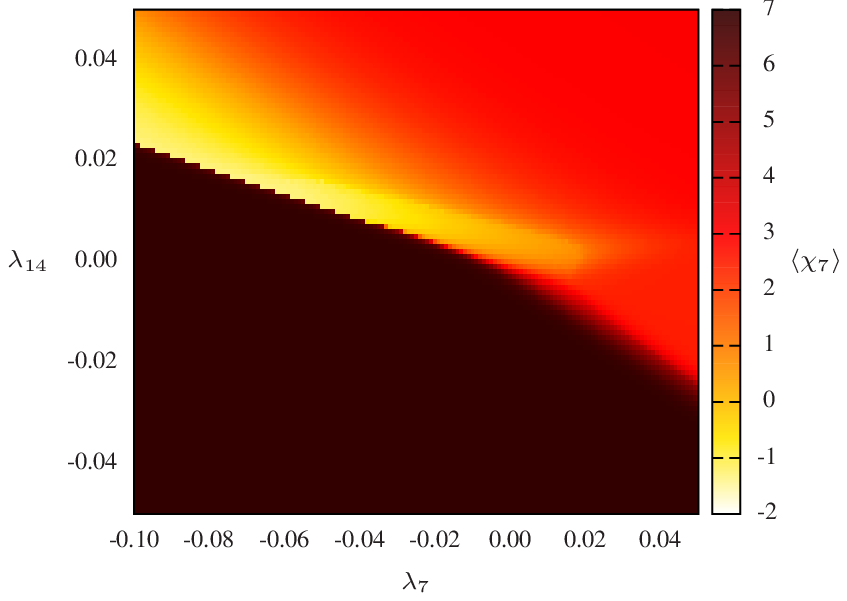}
\caption{Magnetisation $\vev{\chi_7}$ of the discrete $G_2$ Potts model
obtained via MC simulation on an $8^3$ lattice.}
\label{fig:SpinPolyakov}
\end{figure}

The corresponding contour plot as obtained from Monte-Carlo simulations is
shown in  Fig.~\ref{fig:SpinPolyakov}. Again, the mean field approximation and
the Monte-Carlo simulations fully agree over the whole range of coupling constants. Note that the classical behaviour as depicted in 
Fig.~\ref{fig:SpinClassical} can be seen already for rather small coupling constants.

\begin{figure}[t]
\includegraphics{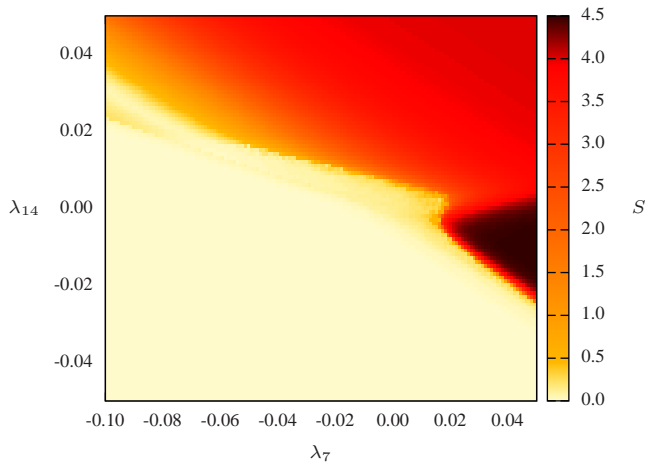}
\caption{Staggered magnetisation $S$ of the discrete $G_2$ Potts model
obtained via MC simulation on an $8^3$ lattice.}
\label{fig:SpinStaggered}
\end{figure}
In order to localize the anti-ferromagnetic phases we also measured the staggered magnetisation
introduced in \eqref{staggered}. The resulting values on a grid in coupling
constant space are plotted in  Fig.~\ref{fig:SpinStaggered}.
In accordance with the classical analysis we detect different anti-ferromagnetic
phases for positive coupling constants in the upper right part of the phase
portrait.

As concerning the relation between the Potts-type model and $G_2$ gluodynamics 
one caveat should be mentioned. In the discrete spin model there exists no real ``symmetric phase''
with a fixed expectation value of $\vc{\chi}$. Even for very weak coupling do the
magnetisations $\vev{\vc{\chi}}$ depend on the couplings. This is a remnant of the
missing centre symmetry of $G_2$. Nevertheless, there exists a first order
phase transition from one (would be symmetric) ferromagnetic phase at
$\lambda_7\approx\lambda_{14}\approx 0$ to a second
ferromagnetic phase with $\trP_\vcx$ directed to the $\mathbbm{1}$-element in $G_2$. 
Even in the simple discrete model we see very pronounced
what happens in real $G_2$ gluodynamics. In the deconfining and confining phase 
there is a non-vanishing Polyakov loop, which still serves as \emph{approximate}
order parameter for confinement since it shows a steep jump at the transition point.


\section{Conclusions}
\label{sect:conclusions}
\noindent
Effective models for confinement with the Polyakov loop as macroscopic degree
of freedom arise naturally from the strong coupling expansion
of $G_2$ gluodynamics. Already the leading order continuous and discrete
effective theories show a rich phase structure with two coexisting phases along transition lines
and three coexisting phases at several triple points. The fundamental 
model with continuous spins and the Potts-type model 
with discrete spins share many properties, although in the absence
of a centre symmetry they need not be in the same universality class. 
The continuous model exhibits a transition from the symmetric to 
the ferromagnetic phase with the same behaviour of the 
Polyakov loop as in $G_2$ gluodynamics, namely a steep jump from
a small (but non-vanishing) Polyakov loop to a loop near the identity 
of $G_2$.

The classical, mean field and Monte-Carlo analysis all lead to a coherent and consistent
picture for both $3$-dimensional effective theories. In particular, the prediction
of the mean field approximation for $\vev{\chi_7}$ and
$\vev{\abs{ \chi_{7,\text{e}}-\chi_{7,\text{o}}}}$ is in excellent agreement
with the corresponding results obtained by detailed Monte-Carlo
simulations. This parallels our findings for $SU(3)$ in \cite{Wozar:2006fi} 
and probably is due to the existence of  tricritical points which lower the upper 
critical dimension in the vicinity of these points.

As concerning the relationship between the continuous effective
models to the underlying $G_2$ gluodynamics we plan to apply 
inverse Monte-Carlo techniques, preferably with demon methods,
to determine the dependence of the coupling in the fundamental model
on the gauge coupling. We hope to present the resulting
curve $\lambda_7(\beta),\lambda_{14}(\beta)$
in a forthcoming publication. However we anticipate that for the critical
Yang-Mills coupling $\beta_\text{c}$ this curve will cross the
transition line between the symmetric and ferromagnetic phases
at small $\lambda_{14}$ and negative $\lambda_7\approx -0.1$ in
Fig.~\ref{fig:effectivePolyakov}.


\begin{acknowledgments}
\noindent
Helpful discussions and earlier collaborations with Tom Heinzl, Tobias
K\"astner,  Kurt Langfeld, and Sebastian Uhl\-mann are gratefully acknowledged.
C. Wozar thanks for the support by the Studienstiftung des
deutschen Volkes. This work has been supported by the
DFG under GRK 1521.
\end{acknowledgments}


\renewcommand{\eprint}[1]{ \href{http://arxiv.org/abs/#1}{[arXiv:#1]}}
\bibliography{literature}

\end{document}